\documentclass[10pt,a4paper]{spie}
\usepackage[utf8]{inputenc}
\usepackage[dvipdf]{graphicx}
\usepackage{makeidx}
\usepackage{color}
\usepackage{multicol}

\title{
    Linearized and Compensated Interferometric System for High-Velocity Traceable Length Calibration on a Metre Scale
    }

\author{
    Simon Rerucha,
	Bretislav Mikel,
	Zdenek Matej,
	Ondrej Herman,
	Miroslava Hola,\\
	Michal Jelinek,
	Petr Jedlicka,
	Ondrej Cip and
	Josef Lazar
\skiplinehalf
Institute of Scientific Instruments of the Czech Academy of Sciences (ISI),\\
 Kr\'{a}lovopolsk\'{a} 147, 612 64 Brno, Czech Republic
\skiplinehalf
}

\authorinfo{Further author information: please send correspondence to e-mail address res@isibrno.cz (S. Rerucha)}

\begin{document}
\maketitle


\begin{abstract}
We report on a traceable calibration system for a 3500mm-long console that carries a 
measurement system for inspecting the diameter of a circular reactor chassis. The system uses two single-pass laser interferometers with homodyne fringe detection for measurement in two degrees of freedom. The hybrid FPGA-microcontroller control module carries out the fringe detection together with the application-specific scale linearization approach and the compensation
of environmental influences such as thermal elongation and the refractive index of air fluctuations.  We demonstrated the system feasibility with an accuracy of a few microns and translation velocity higher than $0.1$ metre per second.

\end{abstract}

%
%
%
%


\section{Introduction}

One of the significant issues within the life extensions of a nuclear power plant reactor is the radiation-induced degradation of the reactor components.\cite{kenik2012_radiationdegradation,zinkle2013challengesnuclear}. Among others, there are concerns that the radiation-induced void swelling could cause changes in the geometry of the reactor chassis, that surrounds the active zone, and that such changes might complicate the operational procedures, e.g. the loading of the reactor with the fuel. In an extreme case, the fuel rods might get stuck in the process, effectively meaning the end of the reactor's operation.

Our work primary aims at a proposed measurement system for periodic assessment of the inner geometry of the reactor active zone chassis. The system consists of two principal components: there is the measurement frame, actually dived into the reactor pool, that carries a battery of short-range displacement sensors and a calibration test-bench that would verify the geometry of the frame itself outside the reactor pool. The geometry is then evaluated by combining these two measurements. 

\begin{figure}[htbp]
	\centering
	a) \includegraphics[width=.57\textwidth]{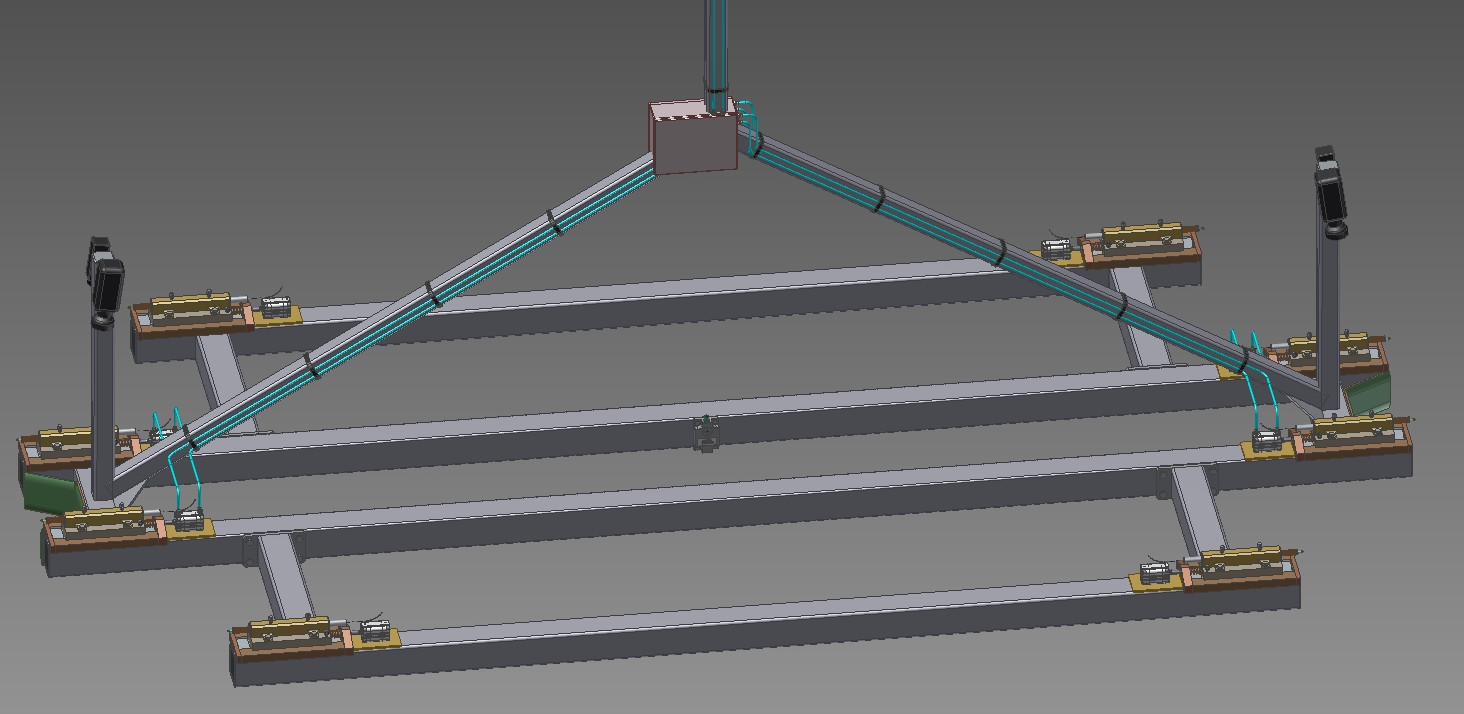}
	b) \includegraphics[width=.28\textwidth]{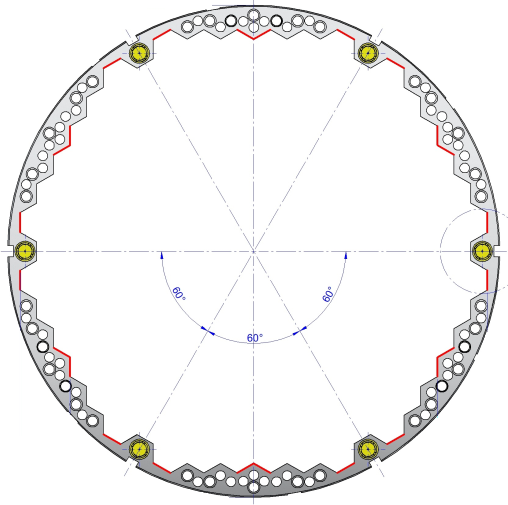}
	\caption{a) the measurement frame, with four parallel measurement axes (A1..A4), that are mounted with a displacement sensors at the tips (B); the reference points (C) on the frame are monitored with a CCD cameras (D); 
			 b) the floor plan of the active zone chassis that shows its hexagonal inner shape and the designated measurement points }
	\label{f_frame}
\end{figure}

The measurement frame, shown in Fig. \ref{f_frame}a), holds four parallel measurement axes. At both ends of each axis, there are displacement gauges with a small range. During the measurement of the pool geometry (sketched in Fig. \ref{f_frame}b),  the measurement frame is immersed into the pool with a crane in three rotations. With each rotation, the frame is consecutively positioned in defined height steps. At each position, the gauges are used to measure the distance to the pool walls at specified measurement points (shown in red in Fig. \ref{f_frame}b). The entire frame is aligned to the guiding pillars of the pool during the course of measurement with the aid of the auxiliary CCD cameras; the horizontal inclination is measured by an inclinometer and the thermal expansion of the entire frame is measured with a specialized fibre Bragg grating sensor.

In this paper, we present and discuss the concept of the interferometric calibration testbench that would allow for a traceable calibration of the geometry of the measurement frame, i.e. for the calibration of the individual axes that are approximately $3300\,$mm long. 
We also incorporate techniques for the compensation of error influences and discuss the (only partially known so far) uncertainty sources so that we can compare the performance with the specified requirements that are as follows: i) total expanded uncertainty with the coverage factor $k = 3$ will be below $100\,\mu$m; ii) measurement will be traceable to the SI metre; iii) the time of a single calibration measurement will be below $1\,$minute. 
As especially the third requirement implies the need for a relatively high bandwidth interferometric phase detection and counting, the signals are processed with a custom dedicated FPGA-based hardware.


\section{Methods}
\label{met}

\subsection{Calibrator Architecture}

\begin{figure}[htbp]
	\centering
	\includegraphics[width=.45\textwidth]{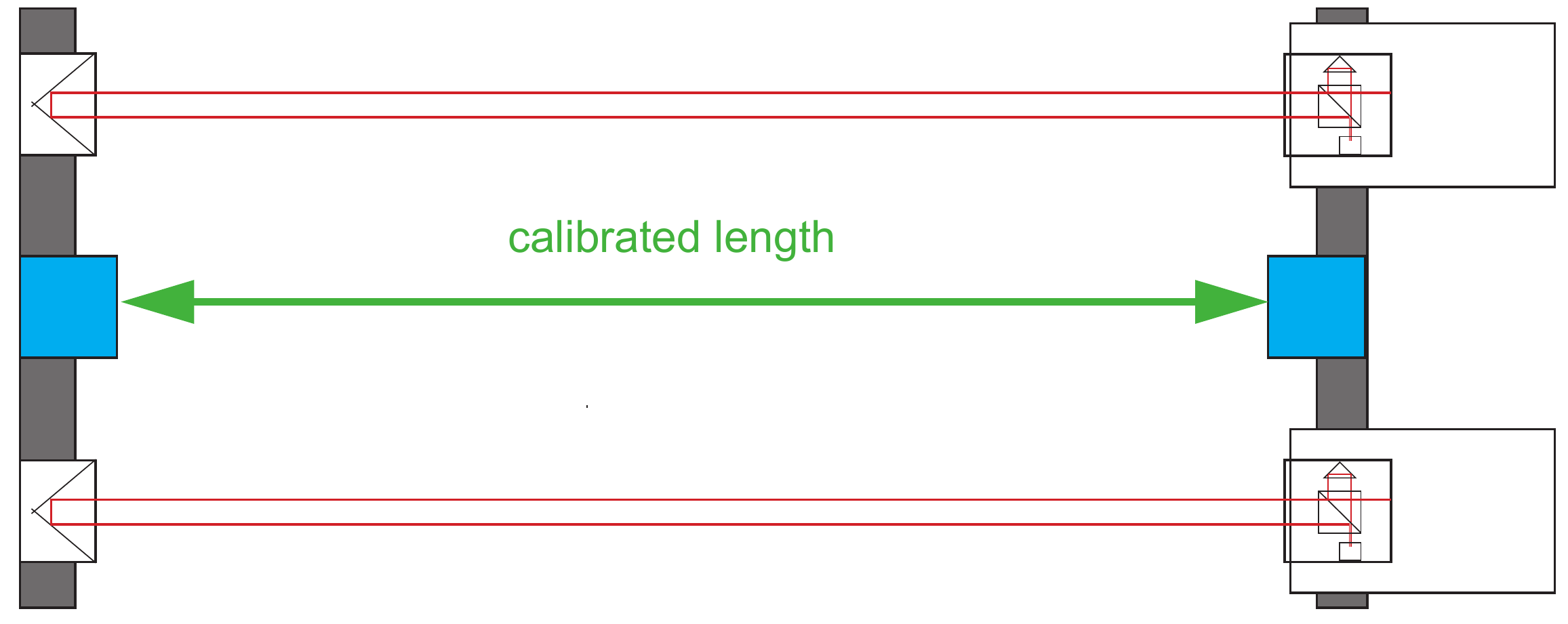}
	
	\caption{Optical arrangement: a pair of Michelson interferometers (or folded Mach-Zehnder) with corner reflectors lies in a single plane with the measurement axis, mitigates the Abbe and cosine errors}
	\label{f_arrangement}
\end{figure}

The calibration test-bench (the calibrator) basically employs a pair of homodyne laser interferometers that measure in two parallel measurement axes, as shown in Figure \ref{f_arrangement}. The two Michelson interferometers use the corner-cube retroreflectors with a polarization beam splitter, one single longitudinal mode He-Ne laser as a source and a four-detector homodyne detection system. 

\begin{figure}[htbp]
	\centering
	a) \includegraphics[width=.35\textwidth]{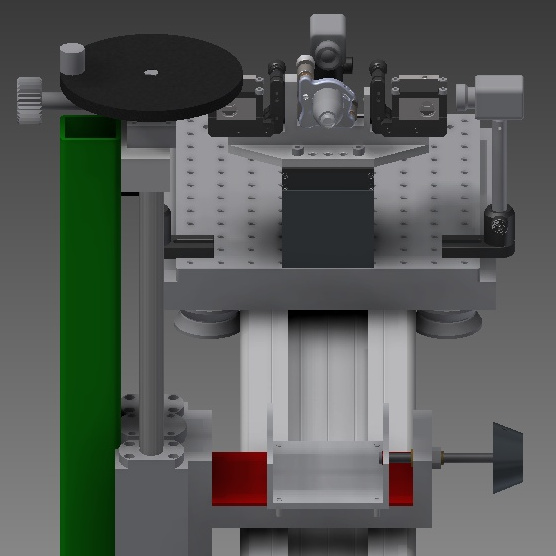}
	b) \includegraphics[width=.35\textwidth]{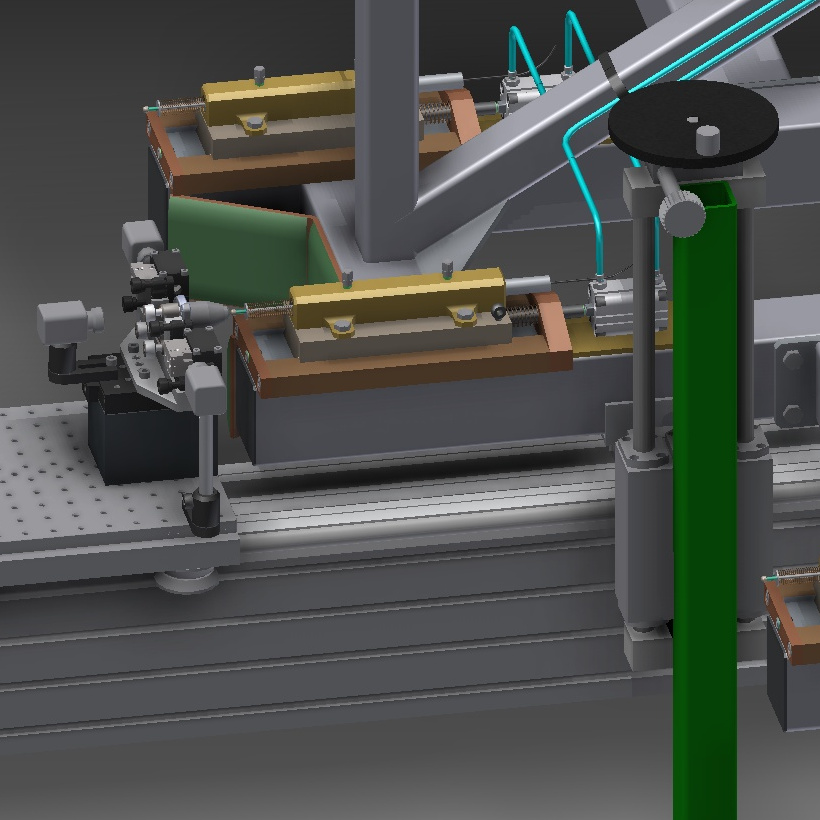}
	\caption{Mechanical construction of the calibrator: a) the movable platform with retroreflectors and the reference contact tips; b) the movable platform with the interferometers and the calibrated frame inserted -- note the displacement gauge aligned to the reference tip}
	\label{f_dwg}
\end{figure}

The measurement axis of the calibrated frame (calibration axis; referred to as the calibrated length in Figure \ref{f_arrangement}) lies in a single plane with the optical paths of the interferometers in between them. The calibration axis is referenced by a pair of contact tips as shown in Figure \ref{f_dwg}a. 

The fixed parts of the interferometers are mounted on a fixed platform and the movable retroreflectors are mounted on the platform of a translation stage.

The course of measurement starts with the stage in position $p_s$ where both the reference surfaces are in contact. Form this position, the stage is moved to a position $p_m$ in which the calibrated frame could be inserted between the reference tips (Figure \ref{f_dwg}b) while the gap between the frame and the reference surfaces still lie within the operational range of the displacement gauges, mounted on the frame axes. Then these gauges are used to measure the distance to the reference surfaces. By subtracting the gauge readings from the difference $p_m - p_s$ we obtain the desired length of the calibrated measurement axis.

\subsection{Error Compensation}

The raw readings from the interferometers are further processed and compensated for error influences: geometry errors, laser fluctuations, interferometer nonlinearities, and fluctuations in environmental conditions.

Regarding the cosine errors and Abbe errors, these are compensated by the optical arrangement that uses two interferometers with their optical paths and the measurement axis lying in a single (horizontal) plane. As for the alignment errors (of the calibrated frame), the calibrator is equipped with three manual translation mechanisms that allow for alignment with an accuracy better than $\pm1\,$mm (with the help of the CCD cameras).

The intensity fluctuations of the laser are compensated by the four-detector arrangement of the interferometer's optical detection unit (e.g. mentioned in Figure 1 in [\cite{Sensors12b}]) that ensure differential reading of the interference phase and consequently a significant mitigation of the offsets in the quadrature signals and the influence of their amplitudes to the phase detection. 

The scale non-linearities, typically induced by imperfections in the optical components, alignment errors in the interferometers' optical paths, guidance errors of the translation stage and systematic offsets in the signal chain, are compensated by an ellipse-fitting technique \cite{Petru1998,CipPetru2000} that ensures linearity better than $0.25\deg$ (i.e. $\lambda / 1440$). There are two principal motivation factors for incorporating the scale linearization: the typical one (but slightly marginal in our case) is the actual improvement in the phase reading accuracy. The other one is that the nonlinearities limit the available bandwidth of the phase reading, which is more crucial for the presented application.

The most significant influence of changing ambient conditions is the fluctuations of the refractive index of air, that influences the effective wavelength of the laser in air. In our system, the temperature, pressure and relative humidity are monitored\cite{Hucl13automatic} and used for a real-time correction of the effective laser wavelength so that the influence is mitigated down to the approximately $10^{-8}$ level.

\subsection{Signal Chain}

\begin{figure}[htbp]
	\centering
	\includegraphics[width=.75\textwidth]{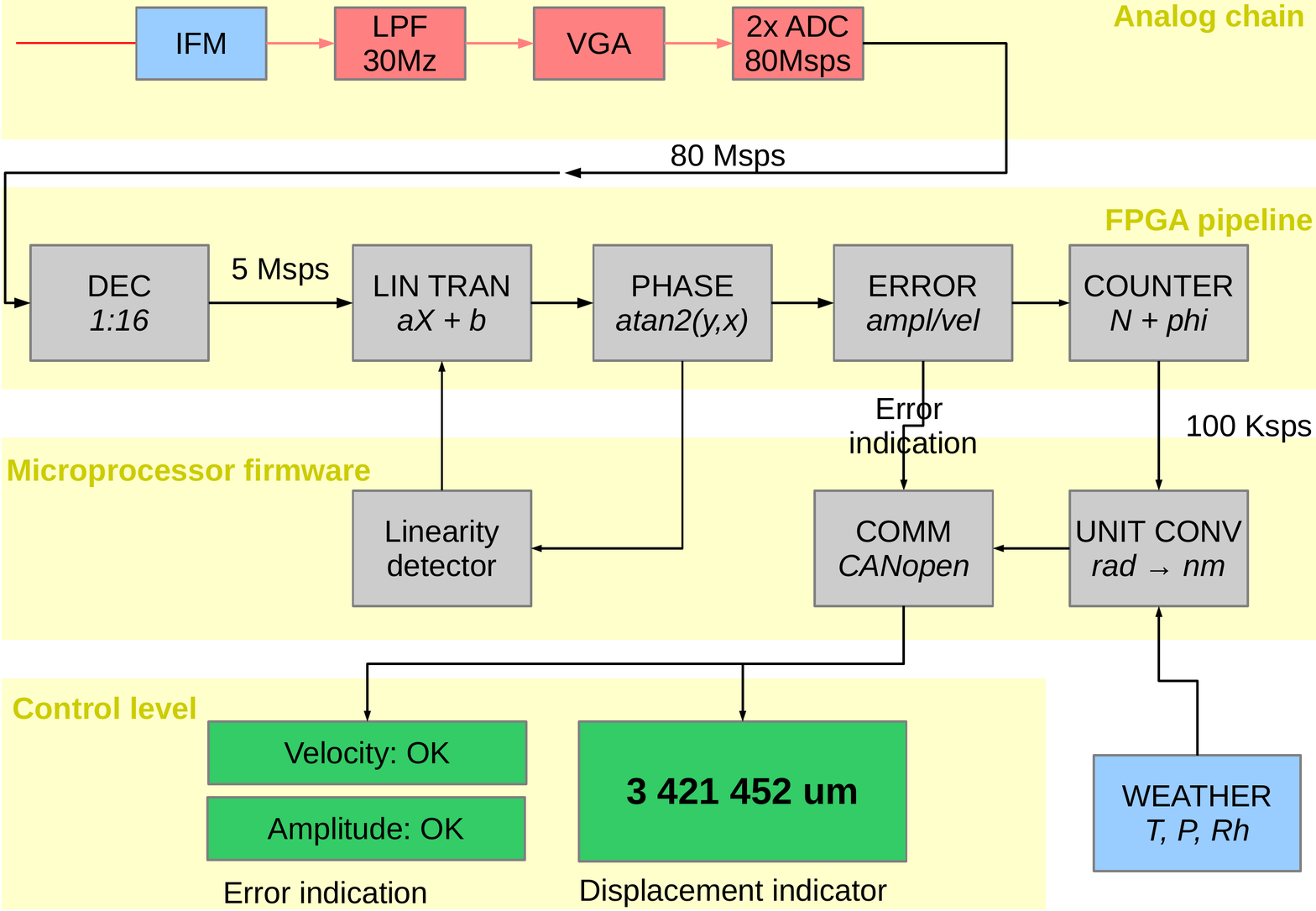}
	\caption{The quadrature outputs from the pair of interferometers are low-pass filtered (LPF), amplified (VGA) to match the range of A/D converters (ADC). The digitised data is decimated (DEC) and linear transformations (LIN TRAN) are applied to correct the scale non-linearities. Then the phase detectors (PHASE) resolve the phase of the interfering beams -- providing also input for dynamic error detection (ERROR) such as velocity and amplitude dropouts -- and the phase is later unwrapped and counted (COUNTER). }
	\label{f_sigblock}
\end{figure}

The entire signal chain, shown in Figure \ref{f_sigblock}, incorporates four processing layers: the analogue chain, FPGA-based pipelined digital processing chain, auxiliary real-time calculations on a microprocessor and a control level, that also incorporate external sensors. The FPGA and microprocessor parts are designed for deployment on a System-on-Chip platform, that combines FPGA and microprocessor core(s) on a single chip.


The analogue chain begins the pair of laser interferometers with output connected to the inputs of the signal module. The signals are low-pass filtered to avoid aliasing, amplified with a variable gain amplifier to match the dynamic range and converted to digital representation with a pair of two-channel analogue-to-digital converters with 16-bit resolution and $80\,$MHz sampling rate. 

In the FPGA pipeline, the digital data are decimated and a linear transformation is applied so that the scale non-linearities are compensated. From the compensated signals the quadrature phase is decoded. From the unwrapped data the fringe overflows and amplitude dropouts are detected before the phase unwrapping.

The subsampled data stream is passed to the microprocessor firmware that ensure following tasks: the phase is converted to displacement; the data are sent to external control software via CANbus and socket; the phase data are used for the fitting part of the scale linearization algorithm and respective parameters are fed to the linear transformation block in the FPGA.


\section{Unit Testing with Simulated Inputs}
\label{exp}

At the current stage of realization, the analogue chain with FPGA pipeline and microprocessor firmware are integrated on a single custom-developed board (signal module) that carries the SoC chip (Xilinx ZYNQ 7020) with peripherals and auxiliary circuitry. We have implemented the necessary firmware parts, that enables us to realize preliminary feasibility tests with simulated inputs, discussed in following subsections.

\subsection{Scale Linearity}

We have implemented the scale linearization method\cite{Petru1998,CipPetru2000} and tested its performance using Monte Carlo simulation-style approach. 

\begin{figure}[htbp]
	\centering
	\includegraphics[width=.90\textwidth]{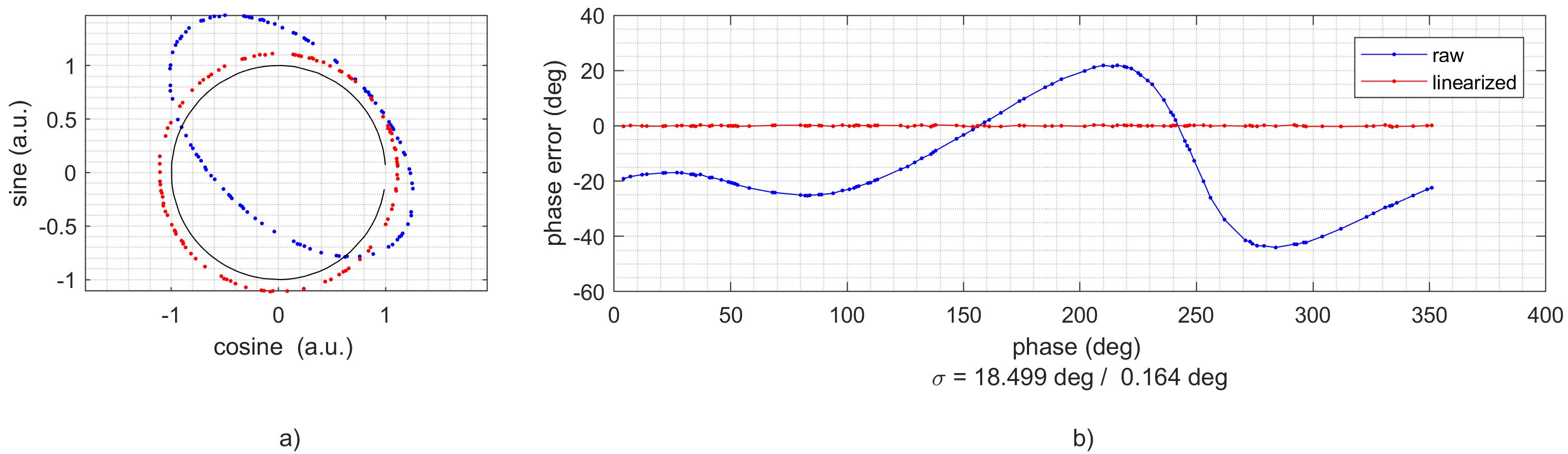}
	
	\caption{The correction of periodic non-linearity in the interferometer: the Lissajous plot (a) with a unit circle in black and periodic linearity (b) before (blue curve) and after the compensation (red curve)}
	\label{f_linsc}
\end{figure}

The simulation input, i.e. vector of pairs $(X_i,Y_i)$ that represent the quadrature ouput from the interferometer, was generated as follows: $n$ points $\phi_i$ representing distinct phase values such that $ 0 \le i \le n; \phi_i \in (-\pi,\pi)$ were randomly generated (with uniform distribution) and from these points the distorted signals were produced as:

\begin{eqnarray*}
X_i &=& K_x \cdot \cos(\phi_i) + x_0 + \epsilon_i. \\
Y_i &=& K_y \cdot \sin(\phi_i + \beta ) + y_0 ) + \epsilon_i.
\end{eqnarray*}

The $K_x, K_y$ denote amplitudes of the quadrature signals, $x_0, y_0$ denote the offsets, $\beta$ denotes an unwanted phase shift from the true quadrature between the two signals and the $\epsilon$ denote a noise component. We used $n = 32\pm7$ points per the phase circle and the parameters were randomly chosen (with uniform distribution) as follows:

\begin{eqnarray*}
	K_x, K_y &\in& (0.75,1.25) \\
	x_0, y_0 &\in& (-0.5,0.5) \\
	beta &\in& (0,\pi / 6) \\
	\epsilon_i &\in& (-0.005,0.005)
\end{eqnarray*}

An illustrative sample of the simulation inputs and corrected output is provided in Figure \ref{f_linsc}.

\begin{figure}[htbp]
	\centering
	\includegraphics[width=.75\textwidth]{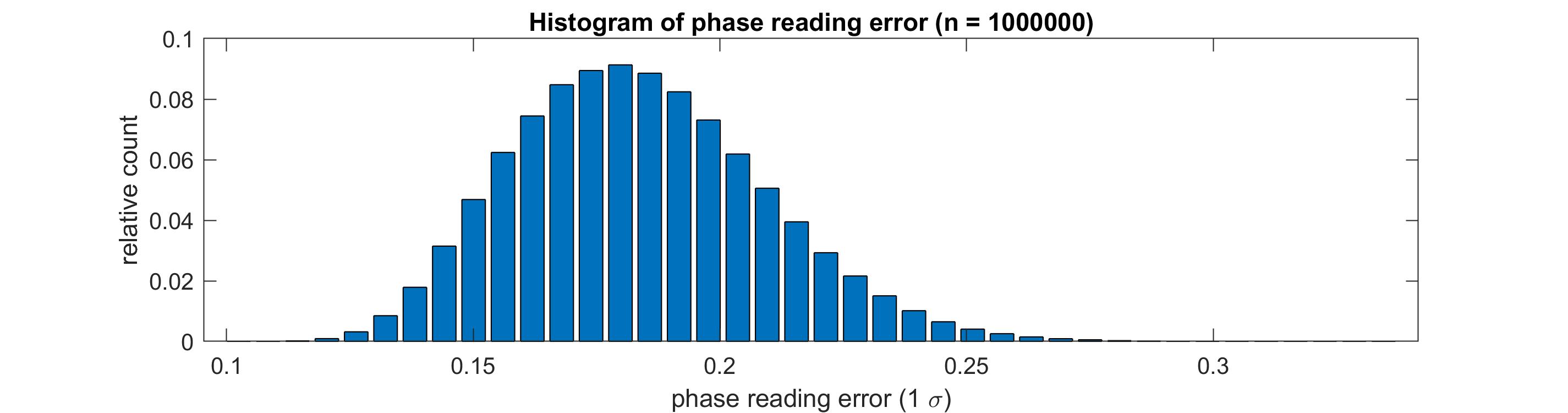}\\
	
	\caption{Histogram of resulting phase reading error after scale linearization}
	\label{f_linscmc}
\end{figure}

This calculation was repeated ($n = 10^6$) and from the resulting phase determination error, shown in the histogram in Figure \ref{f_linscmc},  we have calculated the percentiles that correspond to typical coverage factors and derived the uncertainty of the phase determination as summarised in Table \ref{table_mc}.

 \begin{table}[!t]
	\vskip 2mm
	\renewcommand{\arraystretch}{1.3}
	\caption{Uncertainty of phase determination with the tested scale linearization methods, determined from the simulation results}
	\label{table_mc}
	\centering
	\begin{tabular}{ c c c }
		
		$n$-th percentile	& $k$ 		&  $u_c$ ($\deg$)\\
		\hline
		$68.27\%$	& 1 		&  0.1901 \\
		$95.45\%$	& 2 		&  0.2191 \\
		$99.73\%$	& 3 		&  0.2449 \\
		\hline

	\end{tabular}
\end{table}

\subsection{Signal Chain Performance and Phase Detection Bandwidth}

For the evaluation of the analogue chain and digital pipeline (as displayed in Figure \ref{f_sigblock}), we used a two-channel signal generator (Agilent 33220A) as a signal source. The generator was configured to generate two sinewaves in quadrature, that were connected to the signal inputs of the signal module.

With the 16-bit analogue-to-digital conversion of the interferometers' output the quantization resolution of the phase detection (converted to displacement) is $1.96\,$pm (see Section 2.3 in [\cite{dbr633}] for the calculation), which is far below the target uncertainties of presented application.

\begin{figure}[htbp]
	\centering
	\includegraphics[width=.75\textwidth]{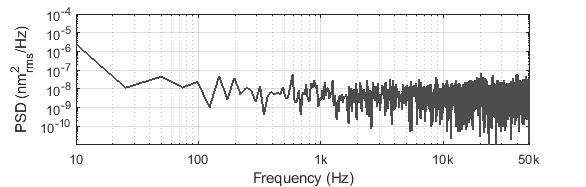}\\

	\caption{The power spectral distribution of the phase reading shows no significant frequency noise components}
	\label{f_psd}
\end{figure}

The reading of the interferometers' signals through the input range exhibited the precision below $0.2\,$mV and the resulting reading of interferometric phase shown precision of $\sigma = 0.019\,\deg$ and exhibited no significant frequency noise components in the power spectra, as indicated in Figure \ref{f_psd}.

\begin{figure}[htbp]
	\centering
	\includegraphics[width=.75\textwidth]{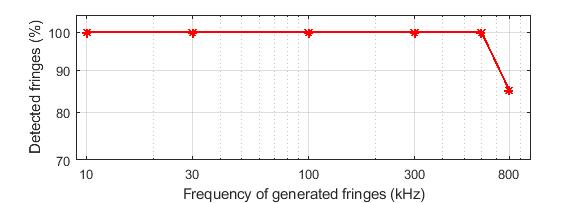}\\
	
	\caption{The results of bandwidth testing indicate that above the fringe frequency of $600\,$kHz the dropouts occur.}
	\label{f_dropouts}
\end{figure}

In order to verify the bandwidth of the fringe counting, we used the generator to produce a specific number of cycles -- so that we simulated a displacement by specified amount of fringes -- with gradually increased frequency of the fringes and we compared the number of fringes on input and the displacement value indicated by the signal module. The test revealed, as shown in Figure \ref{f_dropouts}, that the system detected the displacement up to $0.6 \cdot 10^6$ per second without dropouts.


\section{Discussion and Conclusion}
\label{disc}

The preliminary uncertainty budget summarised in Table \ref{table_uncty} indicates that the major uncertainty source is the nonlinearity of the displacement transducer gauges -- without this influence the combined standard uncertainty would be $0.158\,\mu$m. Please note that the summary does not involve the sources that contribute with less than a nanometre (e. g. the phase detection noise). 

The stability and accuracy of the He-Ne laser wavelength are to be legally calibrated so that the laser will define traceability to the SI metre.

 \begin{table}[!t]
	\vskip 2mm
	\renewcommand{\arraystretch}{1.3}
	\caption{Uncertainty sources and preliminary uncertainty analysis: relative uncertainties calculated for $3.5\,$m displacement where relevant (f. s. -- full scale, DoF -- degrees of freedom)}
	\label{table_uncty}
	\centering
	\begin{tabular}{l c c c c }
		
		Source of uncertainty (type)
		& Value 		& Unit 		& $u_c$ ($\mu$m) 	& notes  \\
		\hline
		HeNe wavelength relative stability (A)
		& $2.10^{-8}$  	& -- 		& $0.0035$			&   \\
		HeNe wavelength absolute accuracy (B)
		& $25$	 		& MHz 		& $0.107$			&   \\
		Scale non-linearity (B)
		& $0.19$	 	& degree	& $0.00017$			&   \\
		Refractive index fluctuation (B)
		& $10^{-8}$	 	& -- 		& $0.0017$			&   \\
		Displacement gauge linearity error (A)
		& $\pm0.5\%$ 	& f.s. 		& $12.5$			&  5mm lift, $2 DoF$ \\
		Frame alignment (B)
		& $\pm1$	 	& mm 		& $0.0825$			&  $2 DoF$ \\
		\hline
		Combined standard uncertainty
		& 			& 			& 17.678			& \\
		Expanded uncertainty for $k=3$
		& 			& 			& 53.035		& \\
		\hline
	\end{tabular}
\end{table}

The test with the simulated output of the interferometer indicates, that with a single-pass interferometer the signal block allow for translation speed up to $0.19\,$m.s$^{-1}$. Considering the range of the calibrator, the translation stage could move along the entire range in less than $20$ seconds. The fact that observed bandwidth is smaller than the theoretical expectation could be  attributed to the nonlinearities in the quadrature signals caused by so far uncompensated offsets in the analogue chain. The figure could be improved by decreasing the decimation ratio at the beginning of the FPGA pipeline (the current FPGA design allow for up to 1:2 decimation ratio). On the other hand, it could be expected that the real signals from the interferometers will also have a negative influence on the bandwidth, e.g. due to additional noise in the signal chain or guidance errors of the large translation stage.

In conclusion, the so far obtained results indicate that the performance meets the specified requirements, while there is still much effort on the system completion, integration and testing to be spent.



\acknowledgments 

The authors acknowledge the support from the Academy of Sciences of the Czech Republic project RVO: 68081731
and, Ministry of Education, Youth and Sports of the Czech Republic (LO1212) together with the European
Commission (ALISI No. CZ.1.05/2.1.00/01.0017); the Ministry of Industry and Trade of the Czech Republic (project FV10336); 
Technology Agency of the Czech Republic (project TE01020233). 


\bibliography{_all}   
\bibliographystyle{spiebib}   

\end{document}